\documentstyle[preprint,aps,epsf,amsmath,amssymb]{revtex}
\tightenlines
\begin{document}
%%%%%%%%%%%% Begin Cover Page %%%%%%%%%%%%%%%%%%%%%%%%%%%%%%%%%%%%%%%%%%
\preprint{
\vbox{
\halign{&##\hfil\cr
	& hep-ph/0105147 \cr
	& ANL-HEP-PR-01-023 \cr
	& UAHEP013 \cr }}
}
\title{Upsilon Decay to a Pair of Bottom Squarks\\}
\author{Edmond L. Berger$^a$ and L. Clavelli$^b$} 
\address{$^a$High Energy Physics Division,
             Argonne National Laboratory \\
             Argonne, IL 60439 \\
		email: berger@anl.gov \\
	$^b$Department of Physics and Astronomy, 
	University of Alabama \\
	Tuscaloosa, AL 35487 \\
	email: lclavell@bama.ua.edu
	}

\date{May 15, 2001}

\maketitle

\begin{abstract} 
We calculate the rate for $\Upsilon$ decay into a pair of bottom 
squarks as a function of the masses of the bottom squark and the 
gluino.  Data from decays of the $\Upsilon$ states could provide 
significant new bounds on the existence and masses of these 
supersymmetric particles.  
\end{abstract} 
\vspace{-0.2in}

%%%%%%%%%%%% End of Cover Page %%%%%%%%%%%%%%%%%%%%%%%%%%%%%%%%%%%%%%%%%

%%%%%%%%%%%%%% Begin Section I %%%%%%%%%%%%%%%%%%%%%%%%%%%%%%%%%%%%%%%%%
{\em Introduction.} The possibility of a light bottom squark, $\widetilde{b}$, 
with mass of order 10 GeV or less, is examined in several 
theoretical~\cite{BHKSTW,CHWW,Nappi,Dreiner,Hou} and 
experimental~\cite{DELPHI,BES,CELLO,CLEO} papers.  If the 
mass $m_{\tilde{b}}$ is less than half 
the mass of one of the $J^{PC} = 1^{--}$ $\Upsilon$ states, then 
the decay $\Upsilon \rightarrow \widetilde{b} \bar{\widetilde{b}}$ might proceed 
with sufficient rate for experimental observation or 
exclusion of a light $\widetilde{b}$.  In this paper, we compute the expected 
rate for $\Upsilon$ decay into a pair of bottom squarks as a function of the masses 
of the bottom squark and the gluino.  The mass of the gluino $\widetilde{g}$ 
enters because the gluino is exchanged in the decay subprocesses we compute. 
The bottom squark and the gluino are the supersymmetric partners of the bottom 
quark and gluon, respectively.  

Measurements at $e^+ e^-$ colliders do not yet constrain significantly the 
existence of $\tilde b$'s in the mass region $m_{\tilde{b}} < 10$ GeV~\cite{BHKSTW}.  
If a light bottom squark ($\widetilde{b}_1$) is an appropriate mixture of left-handed 
($L$) and right-handed ($R$) bottom squarks, its tree-level coupling to the neutral 
gauge boson $Z$ can be made small by adjustment of the $L-R$ mixing, leading to good 
agreement with observables at the $Z$~\cite{CHWW}.  Owing to their $p$-wave 
coupling to the intermediate 
photon and fractional charge $-1/3$, bottom squarks make a small contribution 
to the inclusive cross section for $e^+ e^- \rightarrow$ hadrons~\cite{DELPHI,BES}, 
in comparison to
the contributions from quark production, and $\tilde{b} \bar{\tilde{b}}$
resonances are likely to be impossible to extract from backgrounds
\cite{Nappi}.  The angular distribution of hadronic jets produced in $e^+ e^-$ 
annihilation can be examined in order to bound the contribution of
scalar-quark production.  Spin-1/2 quarks and spin-0 squarks emerge with
different distributions, $(1 \pm {\rm cos}^2 \theta)$, respectively. 
The measured angular distribution~\cite{CELLO} is 
consistent with the production of a single pair of charge-1/3 squarks along with 
five flavors of quark-antiquark pairs~\cite{BHKSTW}.  The exclusion~\cite{CLEO} of 
a $\tilde b$ with mass 3.5 to 4.5 GeV does not 
apply since that analysis focuses on decays into leptons, 
$\tilde b \rightarrow c \em{l} \tilde \nu$ and $\tilde b \rightarrow c {\em l}$.
A long-lived $\widetilde{b}$ or one that decays via baryon-number-violating 
$R$-parity violating couplings would evade the CLEO limitation~\cite{BHKSTW}.  

The possibility that the gluino may be much less massive than most other 
supersymmetric particles is intriguing from different points of 
view~\cite{Fayet,Farrar,LC2,Raby}. The current status of the very low mass gluino 
scenario, further possibilities, and more complete references are found in 
Ref.~\cite{LC2}.  The production of a pair of gluinos with mass in the range 12 
to 16 GeV, decaying with 100\% branching fraction into a bottom quark and a bottom 
squark in the mass range 2 to 5.5 GeV, offers a supersymmetry (SUSY) explanation 
for the large bottom quark cross section at hadron colliders, and it is 
consistent with hadron collider data on the time-averaged $B^0 {\bar{B}}^0$ 
mixing parameter $\bar{\chi}$~\cite{BHKSTW}.  A renormalization group 
argument~\cite{Dreiner} suggests that the existence of a light $\tilde b$ goes 
hand-in-hand with a comparatively light $\tilde g$.   

To compute $\Upsilon$ decay into a pair of bottom squarks, we follow the established 
color singlet approach~\cite{Kuehn,Jones,LC1} in lowest-order perturbative quantum 
chromodynamics (QCD).  We assume that the $\Upsilon$ is a $b \bar{b}$ bound state of 
bottom quarks and that each $b$ carries half of the momentum of the $\Upsilon$ with 
$2 m_b = m_{\Upsilon}$.  Refinements may be considered, but this simple approach is 
sufficient for the purposes of this first paper on the topic.  
 
%%%%%%%%%%%%%% End of Section I %%%%%%%%%%%%%%%%%%%%%%%%%%%%%%%%%%%%%%%%%

%%%%%%%%%%%%%% Begin Section II %%%%%%%%%%%%%%%%%%%%%%%%%%%%%%%%%%%%%%%%

{\em Notation and Normalization.}
Define the four-momentum of the $\Upsilon$ to be $p_{\Upsilon}$.  The 
polarization vector of the $\Upsilon$ is denoted $\epsilon$.  The 
$J^P = 1^-$ bound state of the $b$ and $\bar{b}$ quarks is represented as 
\begin{equation}
\frac{1}{\sqrt 2}\not{\epsilon} \frac{\not{p}_{\Upsilon}+m_{\Upsilon}}{2} A C_{ij}, 
\label{upwavefun}
\end{equation}
where the color matrix $C_{ik} = \frac{1}{\sqrt 3}\delta_{ik}$ expresses the 
color singlet system of the $b$ with color index $i$ and $\bar{b}$ with color index 
$k$.  The coupling strength of the $\Upsilon$ to $b \bar{b}$ is specified in terms of 
an overall constant $A$ which is related to the value of the orbital wave 
function at the origin in momentum space~\cite{TomA}.  This constant may be expressed in 
terms of the electronic width $\Gamma^{\Upsilon}_{\ell \bar{\ell}}$ or the 
hadronic width $\Gamma^{\Upsilon}_{\rm h}$.  In leading order, 

\begin{equation}
\Gamma^{\Upsilon}_{\ell \bar{\ell}} = 
\frac{8 \pi A^2 e^2_{\rm b} \alpha_{\rm em}^2}{m_{\Upsilon}}, 
 \label{eewidth}
\end{equation}
where $e_b = -1/3$.  The lowest-order perturbative QCD expression for a 
three-gluon decay width is~\cite{TomA} 
\begin{equation}
\frac{\Gamma^{\Upsilon}_{\rm h}}{\Gamma^{\Upsilon}_{\ell \bar{\ell}}} = 
\frac{10(\pi^2 -9)}{9\pi} \frac{\alpha_s^3(\mu)}{\alpha_{\rm em}^2}, 
 \label{hadwidth}
\end{equation}
where the strong coupling $\alpha_s(\mu)$ is evaluated at a momentum scale 
$\mu \sim m_{\Upsilon}$.  However, next-to-leading order contributions are 
known to be large~\cite{NLO}.  In our work, we prefer to use the electronic width as 
our source of absolute normalization.  
 
The $\Upsilon$ decays into a pair of bottom squarks $\widetilde{b}$ of 
mass $m_1$ carrying four-momenta $k_1$ and $k_2$ respectively.  This decay 
proceeds via the two-parton to two-sparton subprocess 
\begin{equation}
b + \bar{b} \rightarrow \widetilde{b} + \bar{\widetilde{b}}, 
\label{reaction}
\end{equation}
with exchange of a $t$-channel gluino $\widetilde{g}$ of mass $m_{\tilde{g}}$. 
The subprocess is sketched in Fig.~1.  The subprocess in which the $b \bar{b}$ 
pair annihilates through an intermediate gluon into a $\tilde{b} \bar{\tilde{b}}$
pair is absent because the initial state is a color singlet.  

{\em Mixing.}
The mass eigenstates of the bottom squarks, $\tilde{b}_1$ and $\tilde{b}_2$ 
are mixtures of left-handed (L) and right-handed (R) bottom squarks, 
$\tilde{b}_L$ and $\tilde{b}_R$.  The mixing is expressed as 
\begin{eqnarray}
\tilde{b}_1 = \cos\theta_{\tilde{b}}\tilde{b}_R + 
                                      \sin\theta_{\tilde{b}}\tilde{b}_L, \\
\tilde{b}_2 = -\sin\theta_{\tilde{b}}\tilde{b}_R + 
                                      \cos\theta_{\tilde{b}}\tilde{b}_L.   
\label{mix}
\end{eqnarray}
In our notation, the lighter mass eigenstate is denoted $\tilde{b}_1$.
For the case under consideration, the mixing of the bottom squark is 
determined by the condition that the lighter $\tilde b$ coupling to the 
$Z$ boson be small~\cite{CHWW}, namely $\sin\theta_{\tilde{b}} \simeq 0.38$.

{\em Matrix Element and Decay Width.}
The coupling at the three-point vertex in which a $b$ quark enters and a 
$\tilde{b}_1$ squark emerges (the upper vertex in Fig.~1) is 
\begin{equation}
i g_s \sqrt 2 T^a_{ji} [\cos\theta_{\tilde{b}}P_R - \sin\theta_{\tilde{b}}P_L],
\label{upper}
\end{equation}
where $i$ and $j$ are the color indices of the incident $b$ and final 
$\tilde{b}$, respectively, and $a$ labels the color of the exchanged 
gluino.  $g_s$ is the strong coupling, $\alpha_s = g_s^2/4\pi$.  Here, 
$P_L = (1 - \gamma_5)/2$ and $P_R = (1 + \gamma_5)/2$.  At the lower vertex 
where an antiquark enters and $\bar{\tilde{b}}_1$ emerges, the coupling is 
\begin{equation}
i g_s \sqrt 2 T^a_{k\ell} [\cos\theta_{\tilde{b}}P_L - \sin\theta_{\tilde{b}}P_R],
\label{lower}
\end{equation}
where $k$ and $\ell$ are the color indices of the incident $\bar{b}$ and 
final $\bar{\tilde{b}}$, respectively.  

Define $t = (k_1 - p_{\Upsilon}/2)^2 = - (m^2_{\Upsilon} - 4 m^2_1)/4$.
The invariant amplitude for 
$\Upsilon \rightarrow \widetilde{b}(k_1) + \bar{\widetilde{b}}(k_2)$ is   

\begin{equation}
M = \frac{\sqrt 2 \pi \alpha_s A F_c}{t - m^2_{\tilde{g}}}
Tr\left(\not{\epsilon} (\not{p_{\Upsilon}} + m_{\Upsilon})
[\cos\theta_{\tilde{b}}P_R - \sin\theta_{\tilde{b}}P_L]
[\not{p_{\Upsilon}} - 2 \not{k_1} + 
2 m_{\tilde{g}}]
[\cos\theta_{\tilde{b}}P_L - \sin\theta_{\tilde{b}}P_R]\right).    
 \label{trace1}
\end{equation}
The color factor $F_c = \frac{4}{3\sqrt3} \delta_{jk}$, where $j,k$ label the colors 
of the final squarks.  The relatively large mass of the 
exchanged gluino should justify the use of simple perturbation theory to 
compute the decay amplitude.  Evaluating the trace and using 
$\epsilon \cdot p_{\Upsilon} = 0$, we obtain 
\begin{equation}
M = - \frac{4 \sqrt 2 \pi \alpha_s A m_{\Upsilon} F_c}{t - m^2_{\tilde{g}}} 
\epsilon \cdot k_1.  
 \label{trace2}
\end{equation}
There is no explicit dependence on the mixing angle.  
 
After squaring $M$, averaging over the spins of the $\Upsilon$, and summing over 
the colors of the final bottom squarks, we derive
\begin{equation}
\Sigma |M|^2 = - \frac{32\cdot 16 \pi^2}{3\cdot 9} \alpha_s^2 A^2 
\frac{t m^2_{\Upsilon}}{(t - m^2_{\tilde{g}})^2}.  
 \label{matsq}
\end{equation}
At threshold, $t \rightarrow 0$ and $|M| \rightarrow 0$.  
 
To obtain the decay width of the $\Upsilon$ into $\tilde b \bar{\tilde b}$, we use 
the expression
\begin{equation}
\Gamma = \frac{1}{8\pi}\Sigma|M|^2 \frac{|k|}{m^2_{\Upsilon}}, 
\label{diffwidth}
\end{equation}
where $|k| = \frac{1}{2}\sqrt {m^2_{\Upsilon} - 4m^2_1}$.  
We observe that $d \Gamma  \propto |k|^{2\ell +1}$, with $\ell = 1$, as expected for 
the $p$-wave decay of the $J^P = 1^-$ $\Upsilon$ into a pair of scalars.  
%%%%%%%%%%%%%% End of Section III %%%%%%%%%%%%%%%%%%%%%%%%%%%%%%%%%%%%%

%%%%%%%%%%%%%% Begin Section IV %%%%%%%%%%%%%%%%%%%%%%%%%%%%%%%%%%%%%%%
{\em Predictions.}  Our results are presented in terms of a ratio 
\begin{equation}
R^{\Upsilon}_{\tilde{b} \bar{\tilde{b}}} = 
\frac{\Gamma_{\tilde{b} \bar{\tilde{b}}}}{\Gamma_{\ell \bar{\ell}}} = 
\frac{1}{3} (\frac{\alpha_s(\mu)}{\alpha_{\rm{em}}})^2\frac{m_{\Upsilon}
(m_{\Upsilon}^2-4m_{\tilde{b}}^2)^{3/2}}{(t-m^2_{\tilde{g}})^2}, 
\label{ratio}
\end{equation}
so that the absolute value of the normalization constant $A$ cancels.  
We use the value $\alpha_s(\mu) = 0.22 \pm 0.02$ determined from data on 
$\Upsilon$ decays and evaluated at the scale $\mu = m_b = 4.75$ 
GeV~\cite{Bethke}. This value of $\alpha_s(\mu)$ is 
consistent with determinations from other processes at other physical scales 
when all are plotted in common with the predicted $O(\alpha^4_s)$ QCD 
dependence on $\mu$, with $\alpha_s(M_Z) = 0.1184$ and 
$\Lambda^{(5)}_{\overline{\rm MS}} = 213$ MeV~\cite{Bethke}.  

The choice of a value of $\alpha_s$ smaller than 0.22 might be deemed more 
appropriate.  The presence of a light gluino slows the running of $\alpha_s(\mu)$.  
Extrapolation from $\alpha_s(M_Z)$ to $\alpha_s(m_b)$ with inclusion of a gluino 
of mass 16 GeV reduces $\alpha_s(m_b)$ to $\simeq 0.189$.  The value 
$\alpha_s(9.45 \rm{GeV}) = 0.163 \pm 0.002 \pm 0.014$ is extracted from data on 
radiative decays of the $\Upsilon$~\cite{CLEOrad}.  We comment below on the 
quantitative changes in our results from use of $\alpha_s = 0.16$ rather than 
$\alpha_s = 0.22$.  A smaller value of $\alpha_s$ leads to a more generous 
range of possible values of $m_{\tilde{g}}$ and $m_{\tilde{b}}$.  

As long as $m^2_{\tilde{g}} > |t| = (m^2_{\Upsilon} - 4 m^2_1)/4$, as is true for 
the parameter space of interest to us, the functional dependence of 
Eq.~(\ref{ratio}) shows that the ratio $R^{\Upsilon}_{\tilde{b} \bar{\tilde{b}}}$ 
increases with $m_{\Upsilon}$.  For a given bottom squark mass, larger values of 
$R^{\Upsilon}_{\tilde{b} \bar{\tilde{b}}}$ correspond to smaller values of the 
gluino mass.    

The values of $\Gamma_{\ell \bar{\ell}}(nS)$ listed in the Particle Data 
Group's~\cite{pdg} compilation are $1.32 \pm 0.04 \pm 0.03$ keV, 
$0.52 \pm 0.03 \pm 0.01$ keV, $\sim 0.47$ keV, and $0.248 \pm 0.031$ keV, 
for $n = 1 - 4$, respectively.  The corresponding full widths $\Gamma^{\Upsilon}$ 
are $52.5 \pm 1.8$ keV, $44 \pm 7$ keV, $26.3 \pm 3.5$ keV, and $14 \pm 5$ MeV.
A ratio $R^{\Upsilon}_{\tilde{b} \bar{\tilde{b}}} = 1$ would fall within a 
one standard deviation ($\sigma$) change of $\Gamma^{\Upsilon}$ in all cases, 
and a ratio $R^{\Upsilon}_{\tilde{b} \bar{\tilde{b}}} = 10$ is close to or 
within $1 \sigma$ {\em except} for the $\Upsilon(1S)$ where it corresponds 
to 25\% of $\Gamma^{\Upsilon}(1S)$.  As a working hypothesis, we take 
$R^{\Upsilon}_{\tilde{b} \bar{\tilde{b}}} = 1$ to 10 as a definition of the 
range in which to find solutions for the bottom squark and gluino masses.  

In Fig.~2, we present our results for $R^{\Upsilon}_{\tilde{b} \bar{\tilde{b}}}$ 
as a function of $m_{\tilde{b}}$ and $m_{\tilde{g}}$.  We show sets of curves 
for the choices $R^{\Upsilon}_{\tilde{b} \bar{\tilde{b}}} = 1$ and $10$.  Within 
each set, the curves correspond to $\Upsilon(nS)$ with $n = 1 - 3$, with $n$ 
increasing from the lowest to uppermost curve.  The lines can 
be read as predictions of the width for the corresponding values of bottom 
squark and gluino masses, or as lower limits on the sparticle masses given a known 
bound on $R^{\Upsilon}_{\tilde{b} \bar{\tilde{b}}}$.  

The requirement $R^{\Upsilon}_{\tilde{b} \bar{\tilde{b}}} \le 10$ can be satisfied 
readily within  the range of values of the gluino and bottom squark masses favored 
in Ref.~\cite{BHKSTW}. For example, with 
$R^{\Upsilon}_{\tilde{b} \bar{\tilde{b}}} \le 10$, $m_{\tilde{g}} \le 16$ GeV is 
achieved for $m_{\tilde{b}} \ge 3.5, 3.9, \rm{and} \,4.2$ GeV at the 
$\Upsilon(nS)$ with $n = 1 - 3$.  The more restrictive condition 
$R^{\Upsilon}_{\tilde{b} \bar{\tilde{b}}} \le 1$ at the $\Upsilon(1S)$, along 
with $m_{\tilde{g}} \le 16$ GeV, would demand that $m_{\tilde{b}} \ge 4.5$ GeV, 
still within the range of acceptable solutions in Ref.~\cite{BHKSTW}.  While these 
conclusions are consistent with the range of proposed masses in Ref.~\cite{BHKSTW}, 
our analysis of $\Upsilon$ decays shows nevertheless that bounds on the ratio 
$R^{\Upsilon}_{\tilde{b} \bar{\tilde{b}}}$ are potentially powerful for the 
establishment of {\em lower} bounds on masses.  For example, as is evident in 
Fig.~2, at the $\Upsilon(1S)$ the choice $m_{\tilde{b}} = 3.0$ GeV would require 
$m_{\tilde{g}} \ge 18$ GeV or $\ge 32$ GeV if 
$R^{\Upsilon}_{\tilde{b} \bar{\tilde{b}}} \le 10$ or $\le 1$, respectively.  

Rather than focusing on the value of $R^{\Upsilon}_{\tilde{b} \bar{\tilde{b}}}$, 
we may examine instead the implications of the additional SUSY contribution 
for the full width of the $\Upsilon$.  Because full width of the $\Upsilon(1S)$ 
is the one known most precisely, the most sensitive results on the possible 
exclusion of decays into a pair of $\widetilde{b}$'s with low mass come  
from measurements at the $\Upsilon(1S)$, as long as 
$m_{\Upsilon(1S)} > 2 m_{\tilde{b}}$.  
%The full widths $\Gamma^{\Upsilon}$ of the $\Upsilon(nS)$ states for $n = 1 - 3$ 
%are not measured directly because these states are much narrower than the 
%experimental energy resolution.  The common indirect method is to work from 
%$\Gamma^{\Upsilon} = 
%\Gamma^{\Upsilon}_{\ell \bar{\ell}}/B^{\Upsilon}_{\ell \bar{\ell}}$, where 
%$B^{\Upsilon}_{\ell \bar{\ell}}$ is the leptonic branching fraction.  
The hadronic width of the $\Upsilon$ is calculated in conventional QCD perturbation 
theory~\cite{TomA,NLO} from the three-gluon decay subprocess, 
$\Upsilon \rightarrow 3g$, and,  
as indicated in Eq.~(\ref{hadwidth}), $\Gamma_{3g} \propto \alpha^3_s$.  The SUSY 
subprocess of interest here adds a new term to the hadronic width from 
$\Upsilon \rightarrow \tilde{b} + \bar{\tilde{b}}$.  If this new subprocess is 
present but ignored in the analysis of the hadronic width, the true value of 
$\alpha_s(\mu)$ will be smaller than that extracted from a standard QCD fit by 
the factor $(1 - \Gamma_{\rm{SUSY}}/\Gamma_{3g})^{\frac{1}{3}}$.
For a contribution from the $\tilde{b} \bar{\tilde{b}}$ final state
that is 25\% of $\Gamma^{\Upsilon}(1S)$  ({\em i.e.} 
$R^{\Upsilon}_{\tilde{b} \bar{\tilde{b}}} = 10$), the value of $\alpha_s$ 
extracted via Eq.~(\ref{hadwidth}) will be reduced by a factor of 0.9, at the 
lower edge of the approximately 10\% uncertainty band on the value~\cite{Bethke} 
$\alpha_s = 0.22 \pm 0.02$ used in our calculation.  As remarked above, an even 
smaller value is extracted from data on radiative decays of the 
$\Upsilon$~\cite{CLEOrad}.  From the perspective of this examination of 
$\Gamma^{\Upsilon}$, we deduce that a contribution from $\tilde{b} \bar{\tilde{b}}$ 
decays as large as 25\% of $\Gamma^{\Upsilon}(1S)$ cannot be excluded.  A thorough analysis 
would require the computation of next-to-leading order contributions in SUSY-QCD to 
both the $3g$ and $\tilde{b} \bar{\tilde{b}}$ amplitudes and the appropriate 
evolution of $\alpha_s(\mu)$ with inclusion of a light gluino and a light bottom 
squark.  
 
The data sample is largest at the $\Upsilon(4S)$.  However, for a fixed 
$m_{\tilde{g}} = 14$ GeV, the branching fraction into a pair of bottom squarks 
is about $10^{-3}$ for $m_{\tilde{b}} = 2.5$ GeV, and it drops to about $10^{-4}$ for  
$m_{\tilde{b}} = 4.85$ GeV, below current sensitivity.   

%%%%%%%%%%%%%% End of Section IV %%%%%%%%%%%%%%%%%%%%%%%%%%%%%%%%%%%%%%%

%%%%%%%%%%%%%% Begin Section V %%%%%%%%%%%%%%%%%%%%%%%%%%%%%%%%%%%%%%%%%
{\em Discussion and Conclusions.}
Our study of $\Upsilon$ decay to a pair of bottom squarks demonstrates that 
experimental limits on the ratio $R^{\Upsilon}_{\tilde{b} \bar{\tilde{b}}}$ could 
establish significant new lower bounds on the masses of the gluino and bottom 
squark if $m_{\tilde{b}} < m_{\Upsilon}/2$.  In Ref.~\cite{BHKSTW}, a 
supersymmetry-based contribution is suggested 
as an explanation for the observed size of the bottom quark cross section and 
time-averaged $B^0 {\bar{B}}^0$ mixing probability observed at hadron colliders.  
The proposal requires a bottom squark in the mass range 2.5 to 5 GeV  
together with a gluino in the range 12 to 16 GeV.  In this paper, we show that 
measurements of $R^{\Upsilon}_{\tilde{b} \bar{\tilde{b}}}$ could further 
constrain these allowed regions markedly.  For example, with $m_{\tilde{g}} <
16$ GeV, the very restrictive experimental bound 
$R^{\Upsilon}_{\tilde{b} \bar{\tilde{b}}} \le 1$ at the $\Upsilon(1S)$ requires 
$m_{\tilde{b}} > 4.5$ GeV when $\alpha_s = 0.22$ is used in the computation 
(and $m_{\tilde{b}} > 4.35$ GeV when $\alpha_s = 0.16$ is used).  With 
$m_{\tilde{g}} < 16$ GeV, the bound 
$R^{\Upsilon}_{\tilde{b} \bar{\tilde{b}}} \le 10$ at the $\Upsilon(nS)$ would 
require $m_{\tilde{b}} > 3.5, 3.9$, and 4.2 GeV, respectively, for $n = 1 - 3$ 
and $\alpha_s = 0.22$ (and $m_{\tilde{b}} > 2.6, 3.2$, and 3.5 GeV for $\alpha_s = 
0.16$).  The bound $R^{\Upsilon}_{\tilde{b} \bar{\tilde{b}}} \le 10$ 
implies that the contribution from the $\tilde{b} \bar{\tilde{b}}$ final state
is as much as 25\% of $\Gamma^{\Upsilon}(1S)$, consistent, within uncertainties, 
with the determination of $\alpha_s$ from $\Upsilon$ decays.  
All of these values of $m_{\tilde{b}}$ are acceptable from the perspective 
of Ref.~\cite{BHKSTW} but tend to be towards the upper end of the proposed range. 
Gluinos of mass 12 GeV or less would be incompatible with the restriction 
$R^{\Upsilon}_{\tilde{b} \bar{\tilde{b}}} < 10$ unless 
$m_{\tilde{b}} > 4.2, 4.55$, and 4.75 GeV at the $\Upsilon(nS)$, $n = 1 - 3$ 
(3.9, 4.25, and 4.45 GeV for $\alpha_s = 0.16$).  
Gluinos of mass $> 35$ GeV, or so, are compatible 
($R^{\Upsilon}_{\tilde{b} \bar{\tilde{b}}} \le 1$) with current information on 
the total width of the $\Upsilon$ states for all $m_{\tilde{b}} > 2.5$ GeV.

The presence of a SUSY contribution to the total width $\Gamma^{\Upsilon}$ means 
that the value of $\alpha_s$ extracted from data will be reduced by the factor 
$(1 - \Gamma_{\rm{SUSY}}/\Gamma_{3g})^{\frac{1}{3}}$.  With a smaller $\alpha_s$,  
the contribution from the conventional $3g$ decay is reduced, compensated by 
the SUSY contribution, so as to maintain $\Gamma^{\Upsilon}$.  As remarked, use 
of a smaller value of $\alpha_s(\mu)$ in our calculation allows for a broader range 
of allowed values ({\em i.e.} smaller values) of $m_{\tilde{g}}$ and $m_{\tilde{b}}$.  
Evolution of $\alpha_s(\mu)$ from $\mu = m_b$ to $\mu = m_Z$ is slowed by the presence 
of a light gluino.  Thus, consistency is 
achieved because the smaller $\alpha_s(m_b)$ that allows for a greater contribution 
from $\Upsilon \rightarrow \tilde{b} \bar{\tilde{b}}$ can still be made compatible 
with the same $\alpha_s(m_Z)$.  

A dedicated experimental investigation of the possible decay 
$\Upsilon \rightarrow \widetilde{b} \bar{\widetilde{b}}$ is warranted.    
Direct observation of the decay requires an 
understanding of the ways that the $\widetilde{b}$'s may manifest 
themselves~\cite{BHKSTW}.  
If the $\widetilde{b}$ is relatively stable, the $\widetilde{b}$ could pick 
up a light $\bar{u}$ or $\bar{d}$ and become a $\widetilde{B}^-$ or 
$\widetilde{B}^0$ ``mesino" with $J = 1/2$, the superpartner of the $B$ meson.  
The mesino has baryon number zero but acts like a 
heavy $\bar{p}$ -- perhaps detectable with a time-of-flight apparatus or via 
$dE/dx$ measurements.
R-parity-violating and lepton-number-violating decay of the $\widetilde{b}$ into 
at least one lepton, disfavored by CLEO data~\cite{CLEO}, would imply final 
states with soft leptons.  Possible baryon-number-violating R-parity-violating 
decays of the $\widetilde{b}$ are $\bar{\widetilde{b}} \rightarrow u+s$; 
$\rightarrow c+d$; and $\rightarrow c+s$.  These final states of four light quarks 
might be distinguishable from conventional hadronic final states mediated by the 
three-gluon intermediate state.  Studies of thrust and other event-shape variables 
or of exclusive final states may provide discrimination.  
$\Upsilon$'s produced in $e^{+}e^{-}$ annihilation are polarized transversely 
because of their coupling to the intermediate virtual photon.  Correspondingly, 
the final $\widetilde{b}$ squarks from 
$\Upsilon \rightarrow \widetilde{b} \bar{\widetilde{b}}$ 
will have an angular distribution proportional 
to $\sin^2 \theta$.

A theoretical study similar to that in this paper may be carried out for decays 
of other bottomonium states into pairs of bottom squarks.  

%%%%%%%%%%%%%% End of Section V %%%%%%%%%%%%%%%%%%%%%%%%%%%%%%%%%%%%%%%%

%%%%%%%%%%%%%% Begin Acknowledgment %%%%%%%%%%%%%%%%%%%%%%%%%%%%%%%%%%%%
{\em Acknowledgments.}
ELB is grateful to Tim Tait for clarifying discussions on squark mixing 
and to Brian Harris, David Kaplan, Zack Sullivan, Tim Tait, and Carlos 
Wagner for many valuable conversations.  LC thanks Thomas Gajdosik for 
discussions on angular distributions.  
This research began during a lunchtime conversation at Frontiers in 
Contemporary Physics, Vanderbilt University, March 5 - 10, 2001.
Work in the High Energy Physics Division at Argonne National Laboratory 
is supported by the U.S. Department of Energy, Division of High Energy 
Physics, Contract W-31-109-ENG-38.  LC is partially supported 
under DOE Contract DE-FG92-96ER40967.  
%%%%%%%%%%%%%% End of Acknowledgment %%%%%%%%%%%%%%%%%%%%%%%%%%%%%%%%%%%

%%%%%%%%%%%%%% Begin References %%%%%%%%%%%%%%%%%%%%%%%%%%%%%%%%%%%%%%%%

%%%%%%%%%%%%%% End of References %%%%%%%%%%%%%%%%%%%%%%%%%%%%%%%%%%%%%%%

%%%%%%%%%%%%%% Begin Figure Captions %%%%%%%%%%%%%%%%%%%%%%%%%%%%%%%%%%%
\begin{figure}[tb]
\begin{center}
\epsfxsize= 4.5in %% 6.7in % actual
\leavevmode
\epsfbox{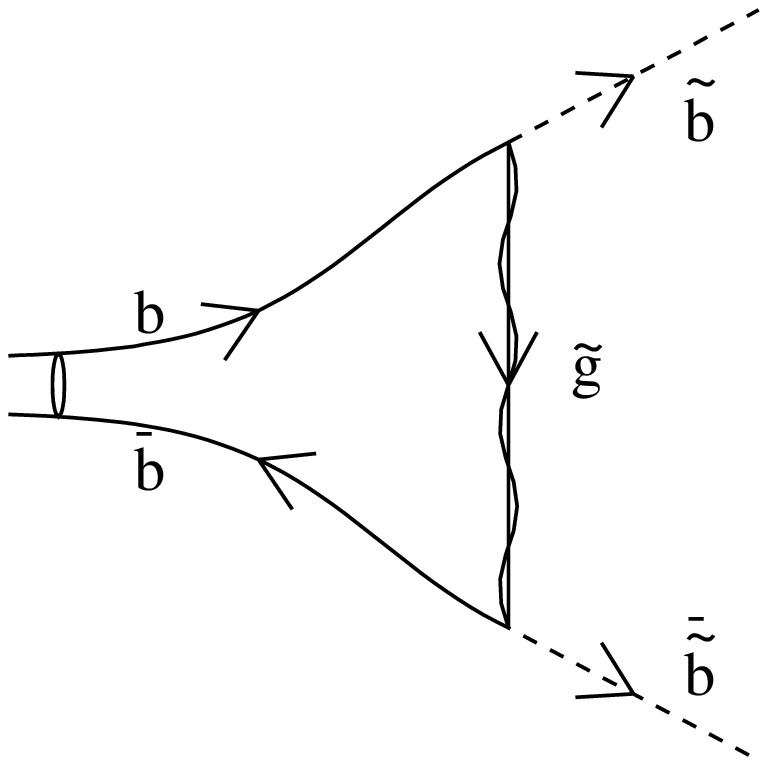}
\end{center}
\caption{Diagram of the decay of a bound $b \bar{b}$ system into a 
$\tilde{b} \bar{\tilde{b}}$ final state, mediated by single gluino 
exchange.}
\label{fig1}
\end{figure}

\begin{figure}[tb]
%\vspace*{-1cm}
\begin{center}
\epsfxsize= 4.5in %% 6.7in % actual
\leavevmode
\epsfbox{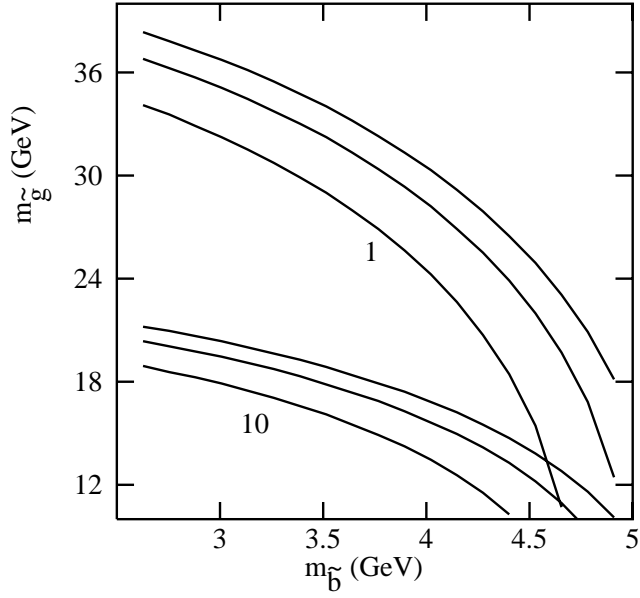}
\end{center}
%\vspace*{cm}
\caption{Loci in the
$m_{\tilde{b}}$ -- $m_{\tilde{g}}$ plane for which the rate for decay into 
a pair of bottom squarks is either 10 times $\Gamma_{\ell \bar{\ell}}$ for each  
$\Upsilon(nS)$, with $n= 1 - 3$ (lower set of curves) or equal to 
$\Gamma_{\ell \bar{\ell}}$ 
(upper set).  Above the curves, the rate would be less.  Within each set 
of curves, the order is (bottom to top) $1S, 2S, 3S$.}
\label{fig2}
\end{figure}

\begin{figure}[tb]
\begin{center}
\epsfxsize= 4.5in %% 6.7in % actual
\leavevmode
\end{center}

\end{figure}

%%%%%%%%%%%%%% End of Figure Captions %%%%%%%%%%%%%%%%%%%%%%%%%%%%%%%%%%

\end{document}